\documentclass[twocolumn,a4,showpacs,preprintnumbers,amsmath,amssymb,nofootinbib]{revtex4-1}
\usepackage{mathptmx, courier, pifont}
\usepackage[scaled=0.92]{helvet}
\usepackage[T1]{fontenc}
\usepackage{textcomp}

\def\bra{\,<\!} \def\ket{\!>\,} 
\usepackage{epsfig}
\usepackage{amssymb}
\usepackage{graphicx}
\usepackage[ps2pdf,colorlinks=true]{hyperref}
\usepackage{textcomp}
\usepackage{color}
\usepackage{colordvi}
\begin{document}
\author{ W.A. Dar$^1$, J.A.~Sheikh$^{1,2}$,  G.H.~Bhat$^{1}$,
R.~Palit$^{3}$ and S.~Frauendorf$^{4}$  }
\address{
$^1$Department of Physics, University of Kashmir, Srinagar,190 006, India \\
$^2$Department of Physics and Astronomy, University of Tennessee, Knoxville, TN 37996, USA\\
$^3$Department of Nuclear and Atomic Physics, Tata Institute of Fundamental Research, Colaba, Mumbai, 400 005,India\\
$^4$Department of Physics, University of Notre Dame, Notre Dame, USA
}

\title{Microscopic study of chiral rotation in  odd-odd A $\sim$ 100 nuclei}

\date{\today}

\begin{abstract}

A systematic study of the doublet bands observed in odd-odd mass $ \sim $ 100 is performed
using the microscopic triaxial projected shell model approach. This mass region has
depicted some novel features which are not observed in other mass regions, for instance,
it has been observed that two chiral bands cross diabatically in $^{106}$Ag. It
is demonstrated that this unique feature  
is due to crossing of the two 2-quasiparticle configurations having different 
intrinsic structures. Further, we provide a complete set of transition probabilities for all 
 the six-isotopes studied in this work and it is shown that the predicted 
transitions are in good agreement with the available experimental data. 

\end{abstract}

\pacs{ 21.60.Cs, 21.10.Hw, 21.10.Ky, 27.50.+e}

\maketitle

Spontaneous symmetry breaking (SSB) mechanism has played a central role to unravel the 
intrinsic structures of the quantum many-body systems. What all can be ascertained
from the experimental data is a set of stationary states that are labeled 
in terms of conserved energy, angular-momentum and parity quantum numbers. The 
underlying mechanism behind 
the observed regularities in quantum spectra is only revealed through the SSB \cite{SF01}. For 
nuclear many-body problem, this has played a pivotal role in classifying the 
observed band structures. Majority of nuclei in the nuclear periodic chart are deformed
with broken rotational symmetry in the intrinsic frame. This broken symmetry is the foundation 
of the popular Nilsson model, which has served an indispensable tool to identify
the intrinsic structures of rotating nuclei with axial symmetry \cite{nilson}. It is known that most of the 
deformed nuclei have axial symmetry with the projection of angular-momentum along the 
symmetry-axis a conserved quantum number. However, nuclei in transitional regions tend to break
the axial symmetry and there are many observations which indicate that nuclei in these regions
have triaxial shapes. The most prominent among these observations, which has attracted a 
considerable attention recently, is the existence of chiral symmetry \cite{SF97,TS04}.
 The occurrence of nuclear chirality
will provide a unique test for the existence of stable triaxial nuclear deformation. 
  
A large number of experimental investigations have been undertaken during the last 
few years to establish the existence of chiral rotation in several  mass regions of the 
nuclear landscape. The candidate chiral bands have been explored and identified in odd-odd and 
odd-mass nuclei in $A \sim 100$ and 130 regions
 \cite{KS01,CV04,TS03,AA01,SU03}. In these regions, chiral geometry is a result 
of the unique orientations of the angular-momenta vectors of protons, neutrons and the core.
For odd-odd nuclei in $A \sim $100 region, odd-proton is a hole-like and odd-neutron is a particle-like
with their angular-momenta along long- and short-axis and the core angular-momentum is directed
along the intermediate axis to minimize the energy. The resulting aplanar total 
angular-momentum can be
arranged into left- or right-handed system, which are related to each other with chiral 
operator -
a combination of time-reversal and rotation by $180^o$. Chiral geometry in the 
laboratory frame manifests itself with the appearance of degenerate doublet bands. 
The properties of the chiral doublet bands have been interpreted using various theoretical approaches
\cite{VD00,PO04,PRM,RPA1,RPA2}.
Although, these models have provided a reasonable description of the properties of chiral
bands, however, there have been some puzzling observations that have 
remained unsolved. For instance,
in $^{106}$Ag, the doublet bands cross diabatically with each other at spin, $I=14$ with moments of inertia of the
two bands being quite different \cite{Ag0106}.
 However, in the neighboring isotopes, there is no such crossing and 
the moments of inertia  of the two bands are similar. The purpose of the present study is to shed 
light on these unsolved issues using the microscopic triaxial projected shell
model (TPSM) approach \cite{JS99}.

It has been demonstrated that TPSM provides an accurate
description of the
high-spin properties of triaxial rotating nuclei. In a recent study, TPSM approach has 
been generalized to multi-quasiparticle configurations and
used to investigate the interplay between the vibrational and the
quasi-particle excitation modes in $^{166-172}$Er \cite{JG11}. It
was demonstrated that a low-lying $K=3$ band observed in these
nuclei, the nature of which had remained unresolved, are built on 
triaxially-deformed two-quasiparticle
states. This band is observed to interact with the $\gamma$-vibrational 
band and becomes favored at high angular-momenta for some Er-nuclei. 
More recently, TPSM approach has been generalized to study odd-odd $^{128}$Cs nucleus in
the mass $\sim$ 130 region \cite{JG12}.

The basic philosophy in the TPSM approach is similar to that of spherical shell model with the only 
exception that deformed basis is employed rather than the spherical one. This allows to investigate
heavier deformed nuclei with a small number of basis states. The basis space of the TPSM approach 
for odd-odd nuclei
is composed of one-neutron and one-proton quasiparticle configurations $:$
\begin{equation}
\{ | \phi_\kappa \ket = {a_\nu}^\dagger {a_\pi}^\dagger | 0 \ket \}  .
\label{intrinsic}
\end{equation}
The above basis space is adequate to describe the
chiral bands in odd-odd nuclei, which are based on one-proton and one-neutron quasiparticle
configurations. The triaxial quasi-particle (qp)-vacuum $ | 0 \ket $ in Eq.~(\ref{intrinsic}) is
determined by diagonalization of the deformed Nilsson 
Hamiltonian and a subsequent 
BCS calculations. This defines the  Nilsson+BCS triaxial qp-basis in the
present work. The number of basis configurations depend on the number of
levels near the respective Fermi levels of protons and neutrons. 
\begin{table}
\caption{The axial deformation parameter ($\beta$) and triaxial
deformation parameter ($\gamma$) employed in the calculation for
Ag-, Rh- and Tc-isotopes. }
\begin{tabular}{c|cccccc}
\hline            & $^{104}$Ag &$^{106}$Ag & $^{104}$Rh & $^{106}$Rh & $^{98}$Tc & $^{100}$Tc\\
\hline
       $\beta$      & 0.149   &  0.158      & 0.202 & 0.237     &0.181      & 0.220\\
       $\gamma$ & 30      &  30         &  30    & 33         & 31       & 34
\\\hline
\end{tabular}
\end{table}

The states $ | \phi_\kappa \ket $ obtained from the deformed Nilsson 
calculations don't conserve rotational symmetry. To  
restore this symmetry, three-dimensional angular-momentum projection technique 
is applied. From each intrinsic state, $\kappa$, in (\ref{intrinsic}) a band is generated through 
projection technique. The interaction 
between different bands with a given spin is taken into account by diagonalising
the shell model Hamiltonian in the projected basis. 
The Hamiltonian used in the present work is
\begin{equation}
\hat{H} = \hat{H_0} - \frac{1}{2} \chi \sum_\mu  \hat{Q}_\mu^\dagger 
\hat{Q}_\mu - G_M \hat{P}^\dagger \hat{P} 
- G_Q \sum_\mu \hat{P}_\mu^\dagger \hat{P}_\mu,
\label{Hamilt}
\end{equation}
and the corresponding triaxial Nilsson Hamiltonian is given by
\begin{equation}
\hat H_N = \hat H_0 - {2 \over 3}\hbar\omega\left\{\beta~\cos\gamma~\hat Q_0
+\beta~\sin\gamma~{{\hat Q_{+2}+\hat Q_{-2}}\over\sqrt{2}}\right\} ,
\label{nilsson}
\end{equation}
where $\hat{H_0}$ is the spherical single-particle shell model Hamiltonian,
which contains the spin-orbit force \cite{Ni69}. The second, third 
and fourth terms in Eq.~(\ref{Hamilt}) represent quadrupole-quadrupole, 
monopole-pairing, and quadrupole-pairing interactions, respectively. 
\begin{figure}[htb]
 \centerline{\includegraphics[trim=0cm 0cm 0cm
0cm,width=0.4\textwidth,clip]{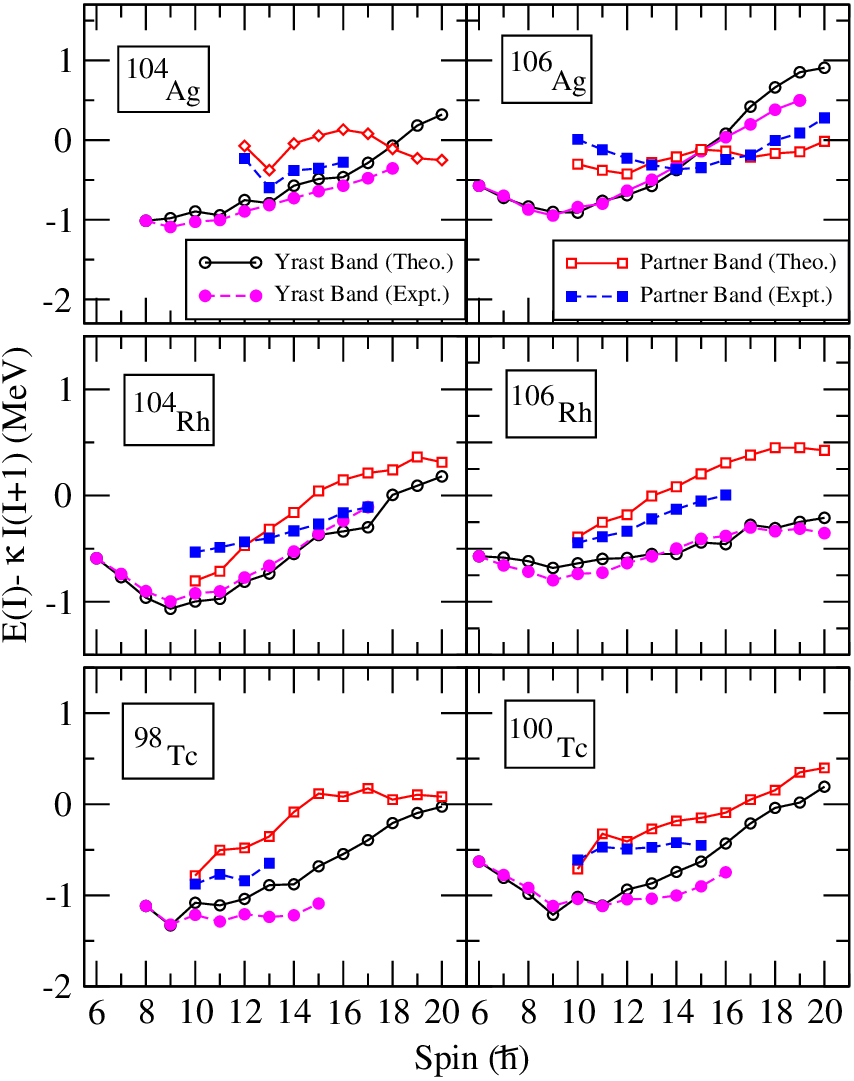}}
\caption{(Color online) Comparison of the measured energy levels of negative parity yrast and
excited bands for $^{104-106}$Rh, $^{104-106}$Ag and $^{98-100}$Tc nuclei the results of TPSM calculations. The value of $\kappa$, shown in y-axis, is defined as $\kappa=32.32 A^{-5/3}$. 
Data are taken from Refs. \cite{Ag0104,Ag0106}, \cite{CV04,106rh} and \cite{98Tc,100Tc} for $^{104-106}$Rh, $^{104-106}$Ag and $^{98-100}$Tc nuclei. }
\label{fig1}
\end{figure}
The axial and triaxial terms of the Nilsson potential in
Eq. \ref{nilsson} contain the parameters $\beta$ and $\gamma$, 
respectively. The strength of the 
quadrupole-quadrupole force $ \chi $ is 
determined in such a way that the employed quadrupole deformation $ \beta $
is same as obtained by the Hartree-Fock-Bogoliubov (HFB) procedure. 
The monopole-pairing force constants $G_M$ used in the calculations are given by
\begin{equation}
G_M ^\nu = \lbrack 20.12 - 13.13 \frac{N-Z}{A} \rbrack A^{-1}, ~~~~~ 
G_M ^\pi = 20.12 A^{-1} . 
\end{equation}
Finally, the quadrupole pairing strength $G_Q$ is assumed to be proportional to
the monopole strength, $G_Q~=~0.16~G_M$. All these interaction strengths are  same as those
used in our earlier study \cite{JG12}. 

\begin{figure}[htb]
 \centerline{\includegraphics[trim=0cm 0cm 0cm
0cm,width=0.3\textwidth,clip]{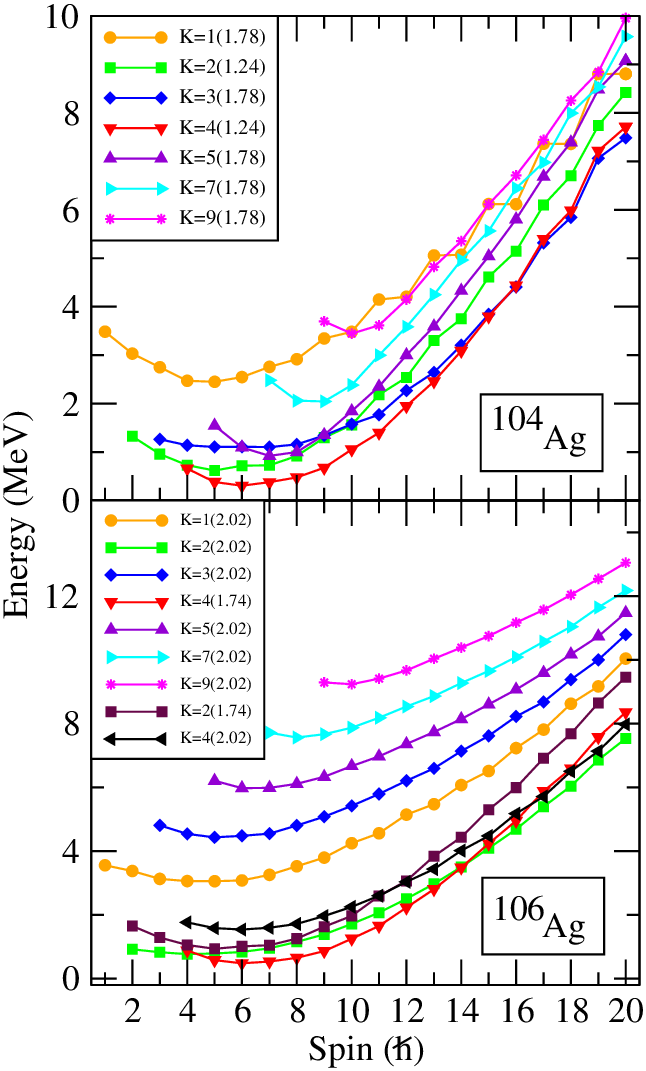}}
\caption{(Color online) Angular-momentum projected bands obtained
 for different intrinsic K-configuration, given in legend box, for $^{104,106}$Ag nuclei. }
\label{fig2}
\end{figure}
\begin{figure}[htb]
 \centerline{\includegraphics[trim=0cm 0cm 0cm
0cm,width=0.3\textwidth,clip]{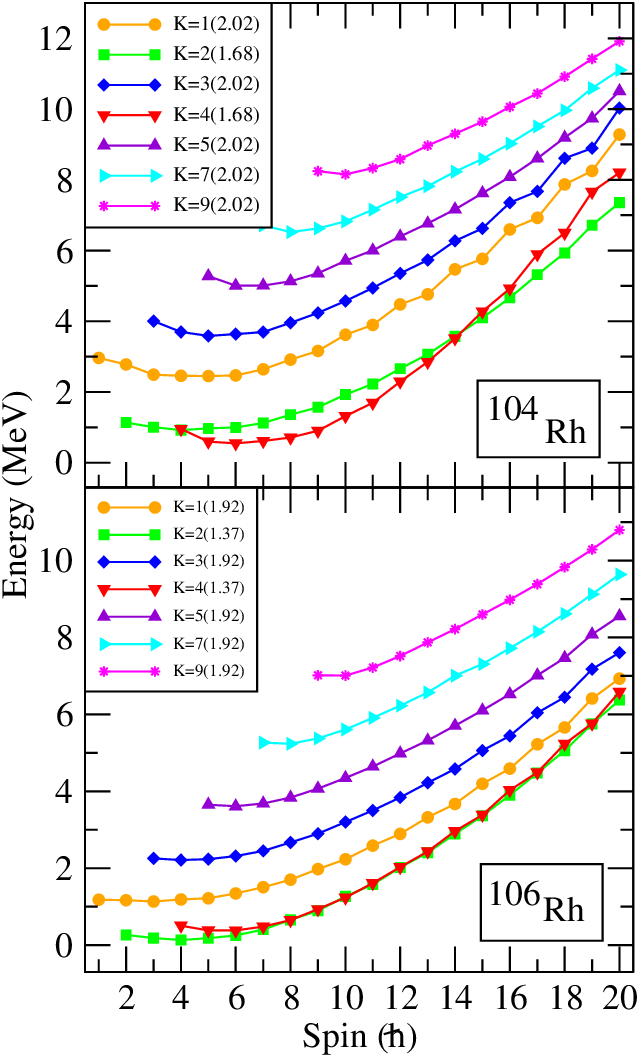}}
\caption{(Color online) Angular-momentum projected bands obtained
 for different intrinsic K-configuration, given in legend box, for $^{104,106}$Rh nuclei. }
\label{fig3}
\end{figure}
\begin{figure}[htb]
 \centerline{\includegraphics[trim=0cm 0cm 0cm
0cm,width=0.3\textwidth,clip]{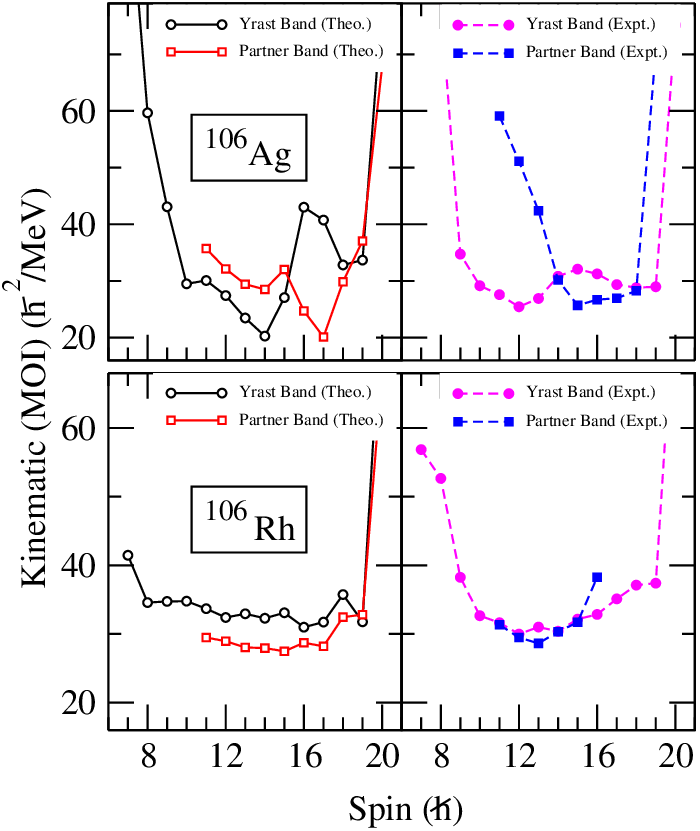}}
\caption{(Color online) Comparison of the kinematic moment of inertia (MOI),
 as a function of spin, obtained from the measured energy levels as well those calculated 
from the TPSM results, for $^{106}$Ag and $^{106}$Rh isotopes. }
\label{fig4}
\end{figure}
\begin{figure}[htb]
 \centerline{\includegraphics[trim=0cm 0cm 0cm
0cm,width=0.3\textwidth,clip]{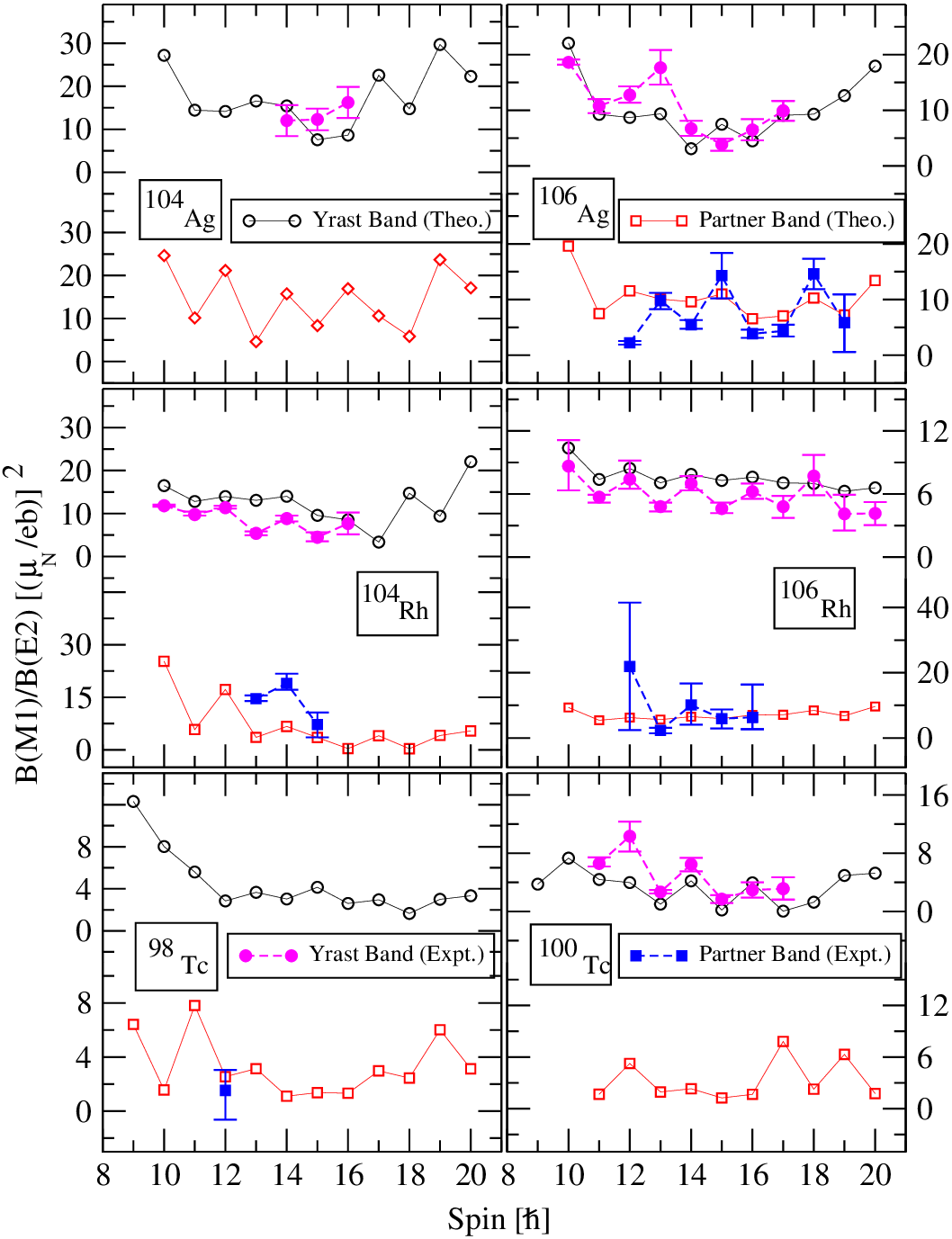}}
\caption{(Color online) Comparison of the experimental and theoretical B(M1)/B(E2) ratios for
$^{104-106}$Rh, $^{104-106}$Ag and $^{98-100}$Tc nuclei. Data are taken from Refs. \cite{Ag0104,Ag0106}, \cite{CV04,106rh},  and \cite{98Tc,100Tc} for $^{104-106}$Rh, $^{104-106}$Ag 
and $^{98-100}$Tc nuclei. }
\label{fig5}
\end{figure}
\begin{figure}[htb]
 \centerline{\includegraphics[trim=0cm 0cm 0cm
0cm,width=0.3\textwidth,clip]{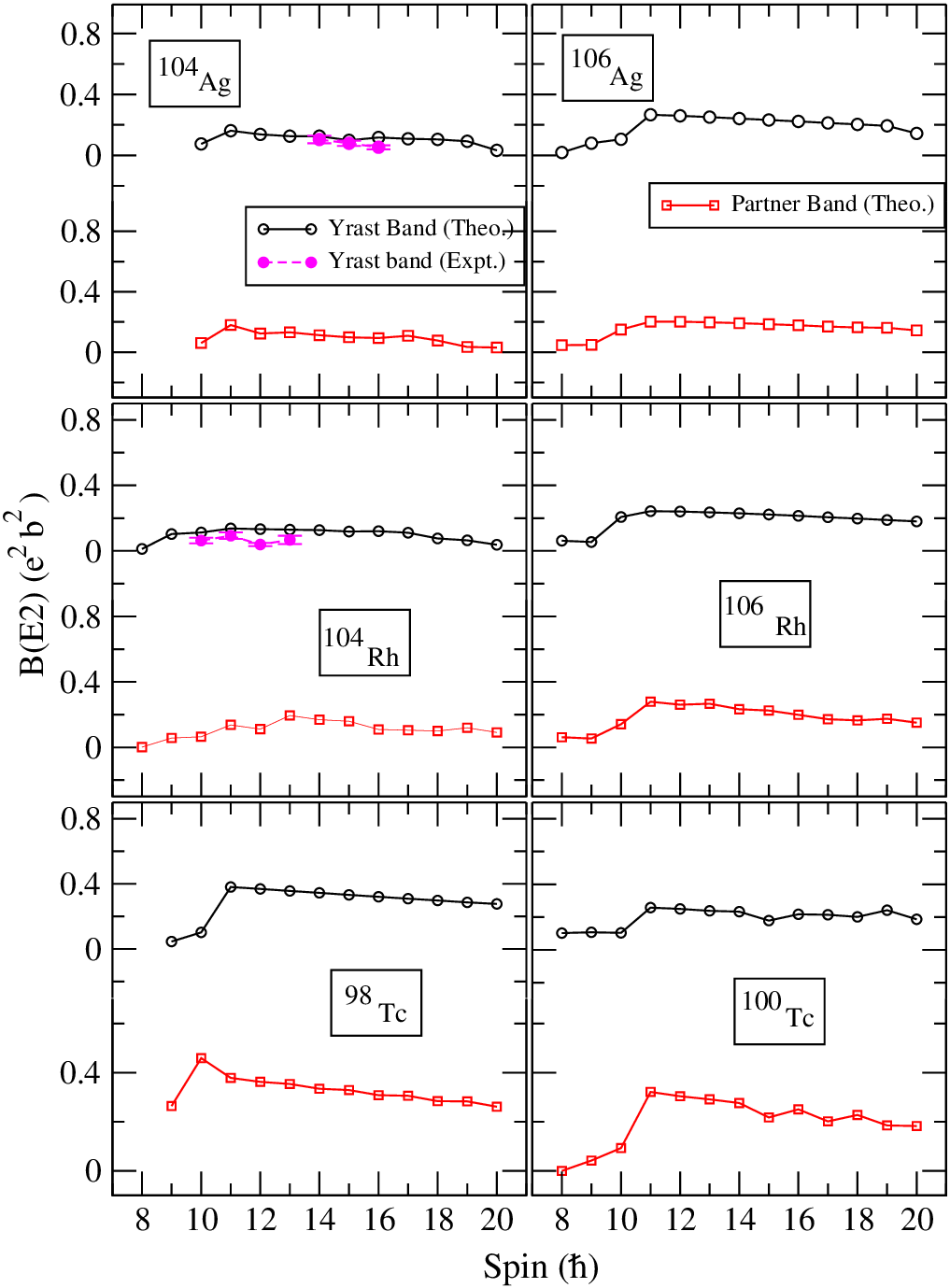}}
\caption{(Color online) Comparison of the experimental and calculated B(E2)
transition strengths for
 $^{104-106}$Ag, $^{104-106}$Rh and $^{98-100}$Tc nuclei. 
Data for $^{104}$Ag and $^{104}$Rh  are taken from Refs. \cite{Ag0104} and
 \cite{TS08}. }
\label{fig6}
\end{figure}
\begin{figure}[htb]
 \centerline{\includegraphics[trim=0cm 0cm 0cm
0cm,width=0.3\textwidth,clip]{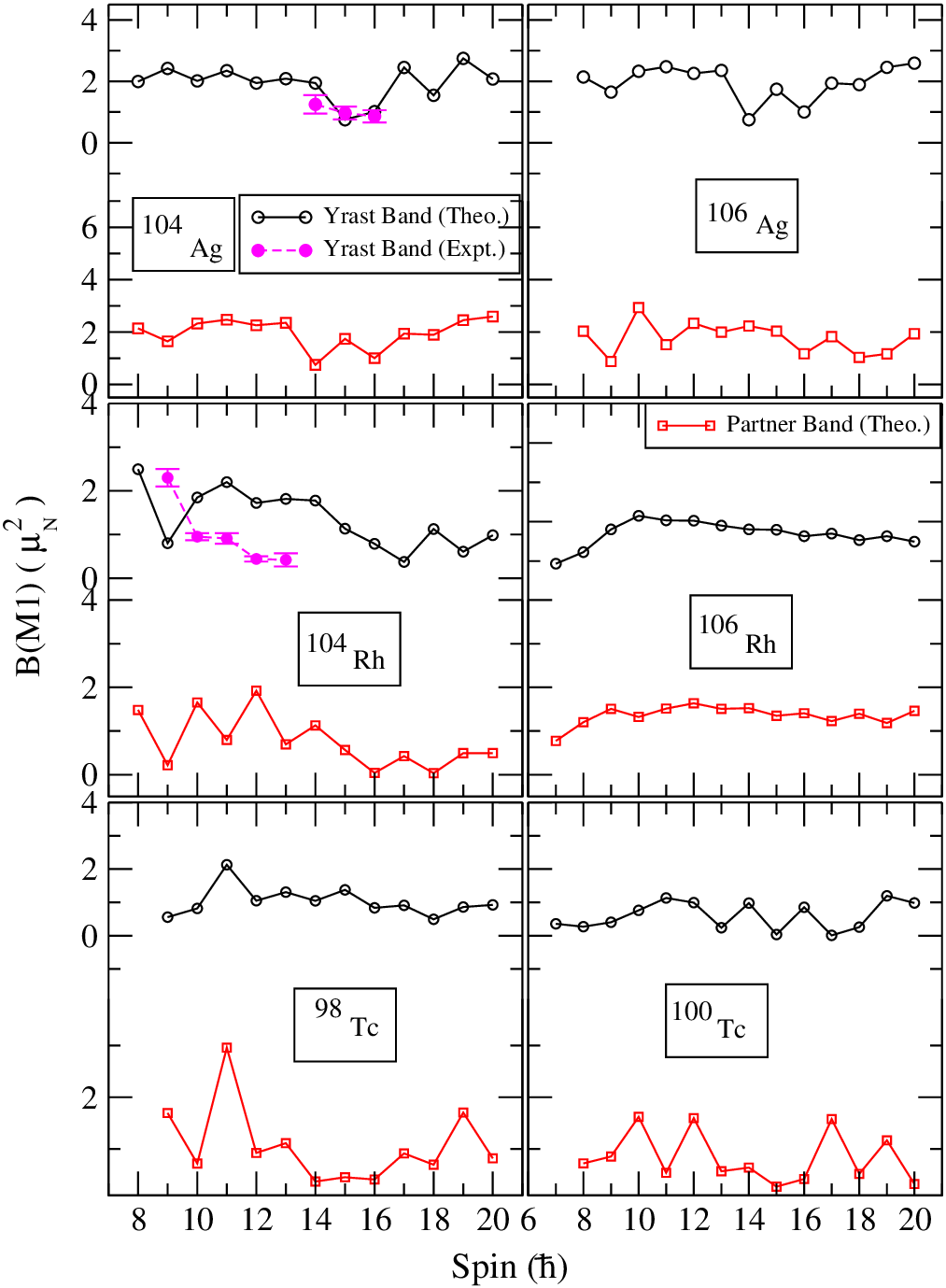}}
\caption{(Color online) Comparison of the experimental and calculated B(M1) 
transition strengths for
$^{104-106}$Rh, $^{104-106}$Ag and $^{98-100}$Tc nuclei. Data for $^{104}$Ag and $^{104}$Rh  are taken from Refs. \cite{Ag0104} and \cite{TS08}.}
\label{fig7}
\end{figure}
\begin{figure}[htb]
 \centerline{\includegraphics[trim=0cm 0cm 0cm
0cm,width=0.3\textwidth,clip]{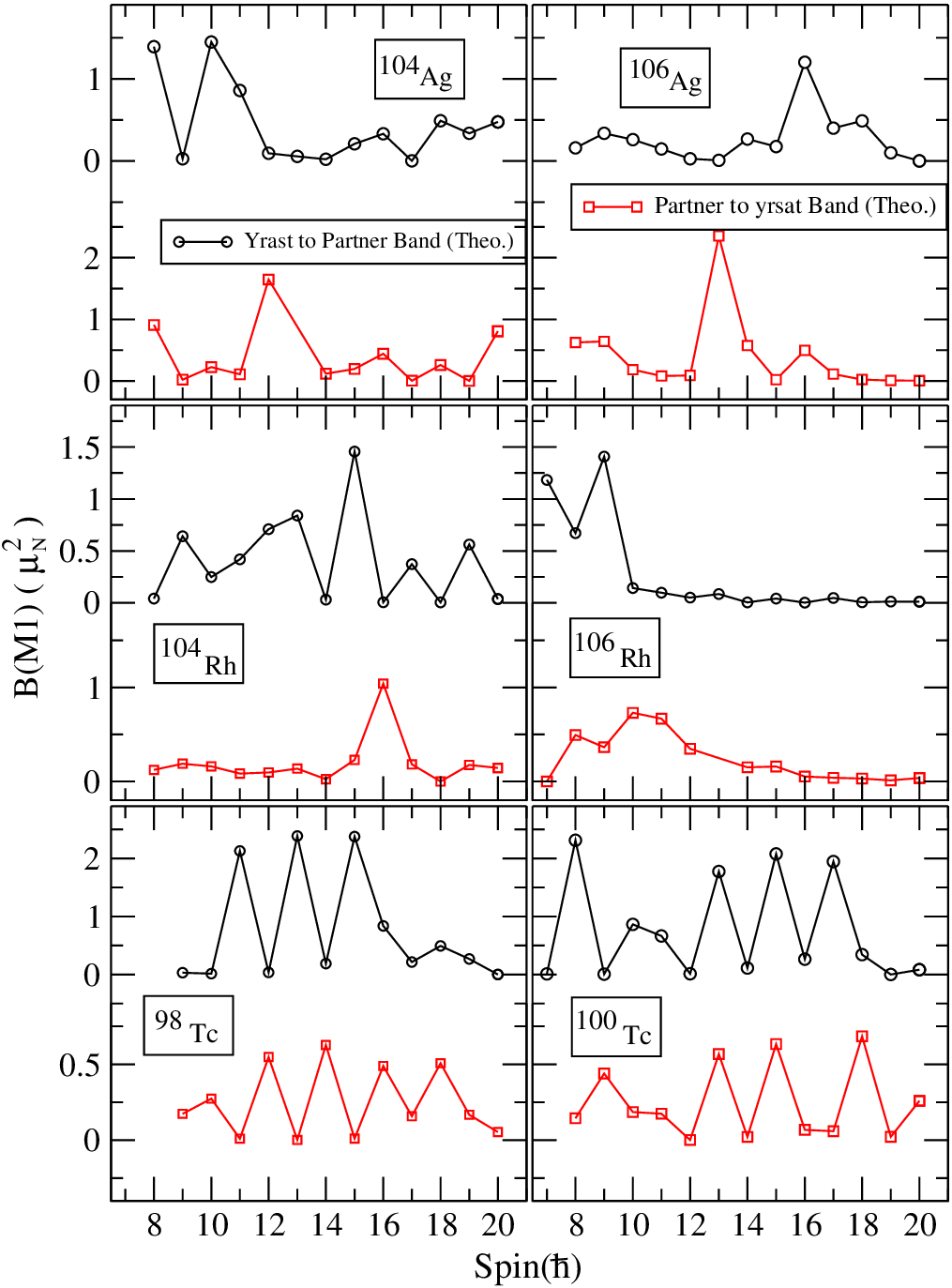}}
\caption{(Color online) Calculated inter-band transitions for
$^{104-106}$Rh, $^{104-106}$Ag and $^{98-100}$Tc nuclei. }
\label{fig8}
\end{figure}
\begin{figure*}[htb]
 \centerline{\includegraphics[trim=0cm 0cm 0cm
0cm,width=0.74\textwidth,clip]{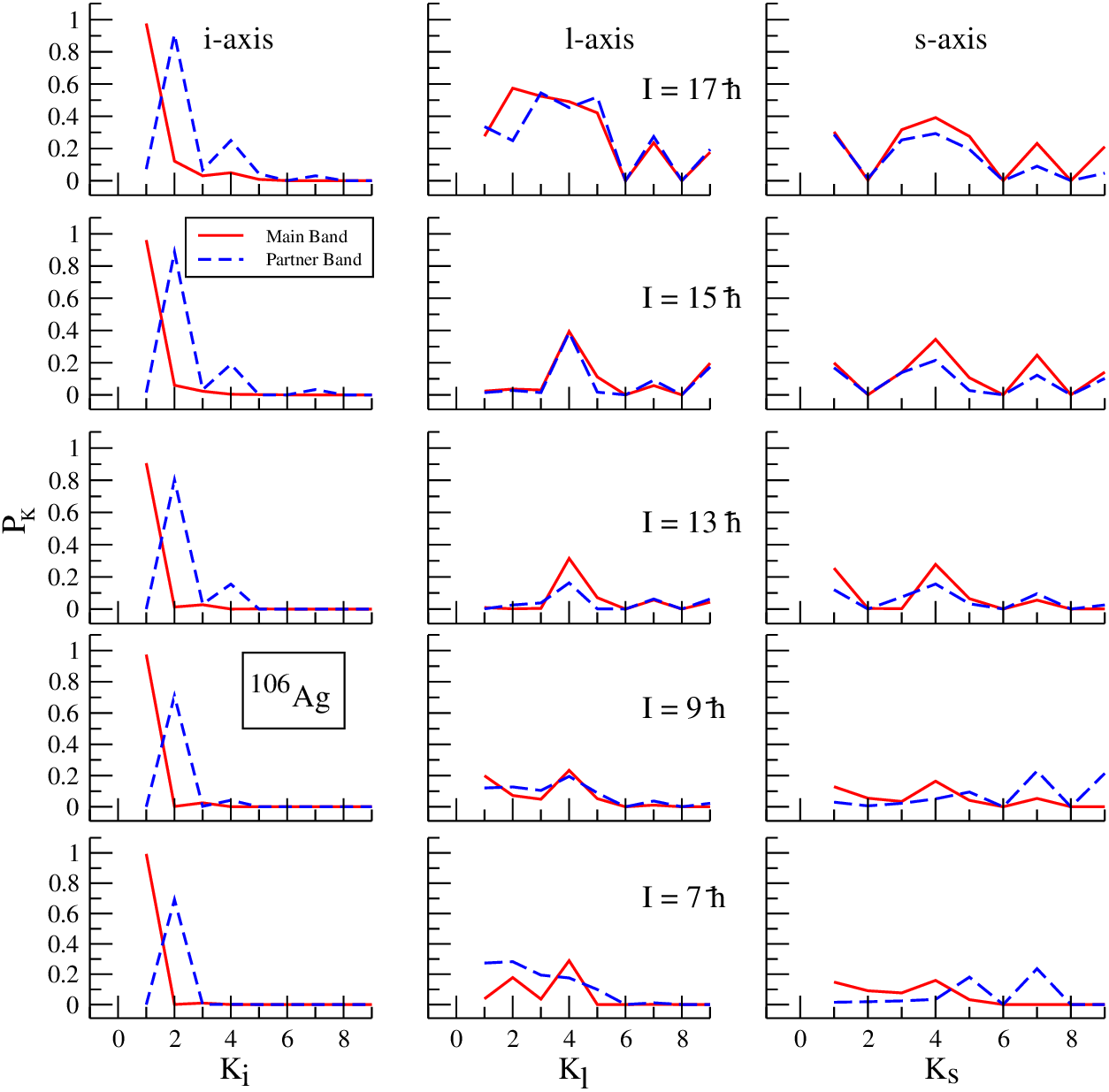}}
\caption{(Color online) The probability distribution for the projection of
total angular momentum, $I^\pi =7^-, 9^-, 13^-, 15^-$ and $17^-$, on the 
intermediate $(i-)$, long $(l-)$, and short $(s-)$ axis for the yrast and its degenerate partner band in $^{106}$Ag  nucleus. }
\label{fig9}
\end{figure*}
\begin{figure*}[htb]
 \centerline{\includegraphics[trim=0cm 0cm 0cm
0cm,width=0.74\textwidth,clip]{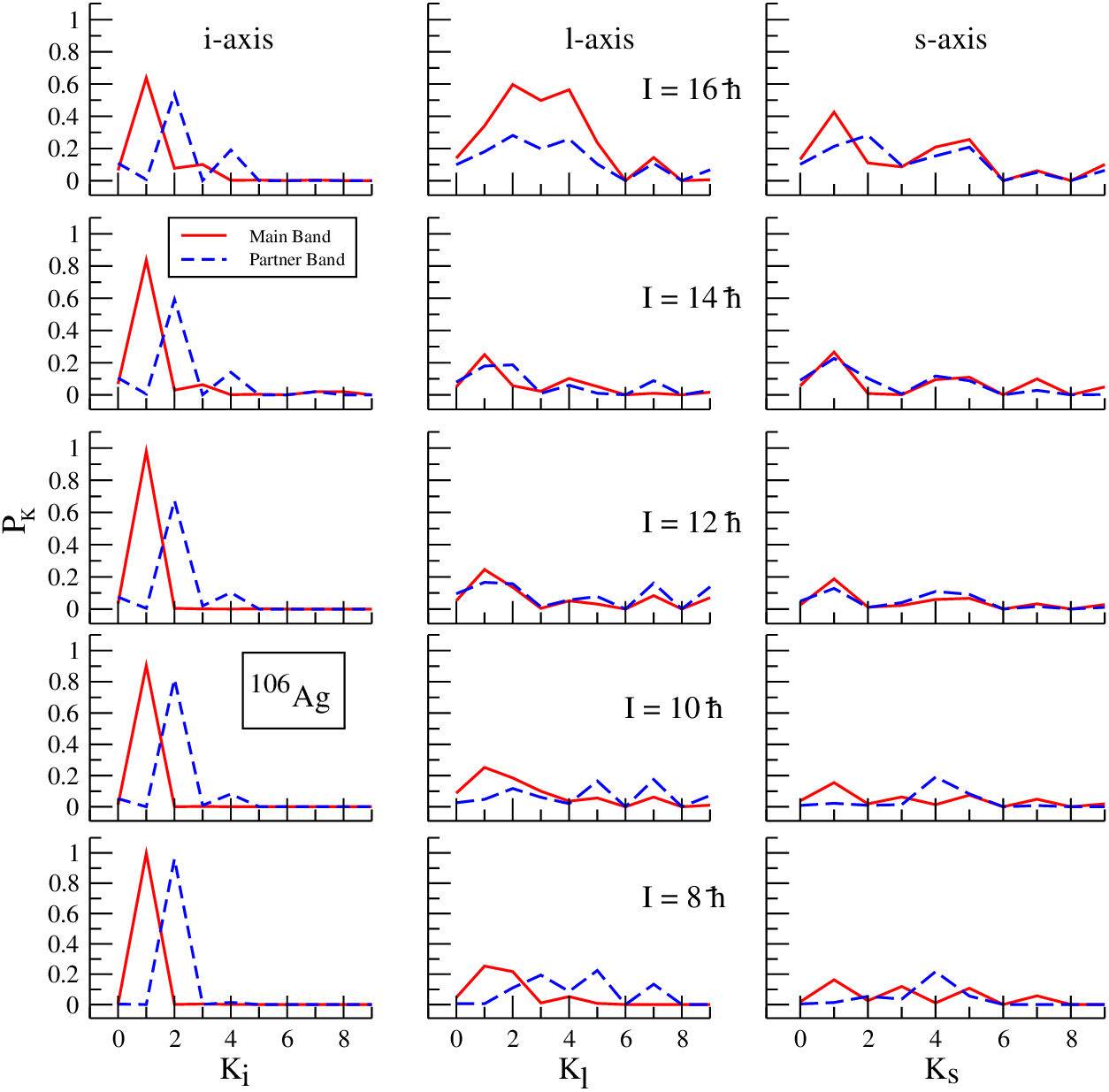}}
\caption{(Color online) The probability distribution for the projection of
total angular momentum, $I^\pi =8^-, 10^-, 12^-, 14^-$ and $16^-$, on the 
intermediate $(i-)$, long $(l-)$, and short $(s-)$ axis for the yrast and its degenerate partner band in $^{106}$Ag  nucleus. }
\label{fig9}
\end{figure*}

In the first stage of TPSM study, the triaxial basis space is constructed by solving
three-dimensional Nilsson potential with deformation parameters, $\beta$ and
$\gamma$. In the present work, the deformation used for the six odd-odd nuclei investigated
are listed in Table 1 and have been adopted from the earlier
studies on these nuclei \cite{Mn95,Ag0104,Ag0106,104rh,106rh,98Tc,100Tc}. The triaxial basis generated are projected onto
good angular-momentum states through three-dimensional 
angular-momentum projection formalism \cite{RS80}.
The projected basis are then employed to diagonalize the shell model Hamiltonian consisting
of pairing plus quadrupole-quadrupole interaction terms. The projected energies, 
obtained after shell
model diagonalization, for the doublet bands in six odd-odd nuclei are depicted and 
compared with the corresponding experimental data in Fig.~1. The energies have been 
subtracted by a core contribution so that the differences become more pronounced. 
It is evident from
the figure that overall agreement between the calculated and the measured energies is quite
remarkable. It is noted from Fig.~1 that theoretical results for $^{104}$Ag and $^{106}$Ag 
depict crossing of the
yrast and the partner bands at I=18 and 15, respectively. The observed partner band in  $^{104}$Ag is known
only till I=16 and it is not possible to verify this crossing. However, for  $^{106}$Ag the 
observed energies are available up to I=20 for both yrast and the partner band and the data clearly
depicts the crossing of the two bands. This crossing of the two bands in $^{106}$Ag has been an
unsolved problem and in order to shed light on the nature of this crossing, the band diagrams
are plotted in Fig.~2 for the two studied Ag-isotopes. 

In band diagrams projected energies, before mixing, with
well defined angular-momentum and the projection ($K$), 
obtained from the quasiparticle-configurations close to the Fermi surface, are plotted. These
diagrams are quite instructive and provide a unique information on the intrinsic structures of the 
projected bands. It is noted from Fig.~2 that the lowest band in $^{104}$Ag has K=4  up to I=17
and then this band is crossed by another band having K=3, which originates from a different
quasiparticle configuration. This figure also provides us a possible explanation as to why 
the partner band is not observed below I=12 as it is noted that for low-spin states, K=3
band is higher in energy and becomes unfavored. Of course, in the final analysis, the 
situation will be somewhat different due to band mixing, but we do expect that this overall
behavior of the bands shall prevail.
For $^{106}$Ag, the lowest band in bottom panel of Fig.~2 has K=4 till I=14 and then
this band is crossed by K=2 band, which again originates from a different quasiparticle
configuration as in $^{104}$Ag. Therefore, the reason for the observed diabatic crossing between
the yrast and the partner bands in $^{106}$Ag is due to crossing of two states having different
intrinsic quasiparticle configurations. In the final results after diagonalization,
the bands are mixed at the crossing due to degeneracy, but below and above the crossing
point, the two bands retain their individual configuration. It was, as matter of fact,
conjectured in the experimental paper \cite{Ag0106} that yrast and partner bands in
$^{106}$Ag may have different quasiparticle configurations and the present theoretical
analysis confirms the postulation.
 
In Fig.~3, the band diagrams for two studied isotopes of $^{104}$Rh and $^{106}$Rh
are displayed and it is quite interesting to note that both these isotopes
depict crossing between the lowest two bands as for the Ag-isotopes. However, 
the important difference is that, in contrast to the Ag-isotopes,
Rh-isotopes have lowest two bands belonging to the
same intrinsic configuration and projected to different K-states. The lowest two 
bands mix strongly due to their identical nature and consequently the observed
yrast and the parnter bands don't depict diabatic crossings in the Rh-isotopes. In
order to elucidate that the structures of the doublet bands in Ag- and Rh-isotopes
are quite different, even after mixing, the moments of inertia of  $^{106}$Ag 
and $^{106}$Rh, obtained
from the projected energies after diagonalization, are displayed
in Fig.~4 as illustrative examples. It is quite evident from this figure 
that moments of inertia of  the yrast and the partner bands for $^{106}$Rh
are similar, both theoretical and experimental, but for $^{106}$Ag are quite different.
The band diagrams for the two studied Tc-isotopes don't depict any band crossings and are not
discussed here.

We have also evaluated transition probabilities using the projected wavefunctions 
after diagonalization with the expressions given in ref.~\cite{JG12}. The parameters
of 
$g_l^\pi = 1, ~
g_l^\nu = 0, ~   
g_s^\pi =  5.586 \times 0.85 ~ $,~$ g_s^\nu = -3.826 \times 0.85 $
and the effective charges of $e^{\pi}=1.5e$ and $e^{\nu}=0.5e$ have been employed as
in our earlier work \cite{JG12}.
The calculated transition probabilities are provided in Figs.~5 to 8. In Fig.~5,
B(M1)/B(E2) ratios of the yrast and the partner bands are displayed for all the 
studied isotopes. The calculated ratios are noted to be in good  agreement with the 
available experimental data. B(E2) and B(M1) transitions obtained from the TPSM study are separately
shown in Figs.~6 and 7. B(E2) transitions for $^{104}$Ag depict a constant behavior 
as a function spin and for $^{106}$Ag there is an increasing trend at low-spin, but for higher
spins, B(E2)'s are again constant. B(E2) transitions for $^{104}$Rh and  $^{106}$Rh have similar
behavior as those of corresponding Ag-isotopes.  Both $^{98}$Tc and  $^{100}$Tc, show
increasing trend in B(E2) for low spins and for higher spins, constant behavior is again
observed. 
The in-band B(M1) transition probabilities, shown in Fig.~7,  depict odd-even staggering
as expected for chiral geometry. The available experimental data, also
shown in Fig.~7, demonstrate a nice agreement for $^{104}$Ag 
but discrepancies are noted for $^{104}$Rh. More data is clearly required to probe the
limitations of the TPSM approach. The calculated inter-band transitions are depicted in Fig.~8 and
have opposite phase to those of in-band transition of Fig.~7, as expected for the
chiral bands. It has been demonstrated in a simple model study that some special selection
rules are obeyed by the transition probabilities in the chiral limit \cite{TS04}, 
in particular, the in-band B(M1) transitions have opposite phase to that of inter-band
transitions. Although, these selection rules are not exactly applicable to 
the present realistic model study, it is expected that these rules be 
approximately satisfied.

In order to probe  the chirality of the nuclei studied in the present work,
we have evaluated the K-distributions of the wavefunctions for the doublet bands and as
an illustrative example, the results are presented for $^{106}$Ag. 
The projections along the quantization axis of
intermediate, $i-$ and short, $s-$ are simply obtained by using the $\gamma$-values of $90^0$ and 
$150^0$ \cite{qi09}. In the present TPSM approach, as compared to the particle-rotor,
the projected basis are not orthogonal and in Figs. 9 and 10 only the diagonal
components of the expression :
\begin{equation}
P(K,K^\prime) = \sum _{\nu\nu^\prime} c_{K\nu} \bra \nu | P^I_{KK^\prime} | \nu^\prime \ket c_{K\nu^\prime}
\end{equation}
are shown. In the above equation, $c_{K\nu}$ are the amplitudes of the yrast- and the 
partner-band wavefunctions.

Figs. 9 and 10 provide the diagonal K-distributions  $P_K=P(K,K)$ for the projection of 
the total angular momentum along the intermediate $(i)-$, long $(l-)$ and 
short $(s-)$ quantization axis for odd- and even- angular-momentum states, 
respectively. 
We interpret $P_K$ as a measure for the probability of the projection of the total 
angular momentum on the respective axis and its 
distributions in the following manner. The g$_{9/2}$ proton hole tends to align its $\vec j$
with the l-axis and the h$_{11/2}$ neutron particle  tends to align its $\vec j$
with the s-axis. The projections of $\vec j$ of the two quasiparticles generate part of
$K_l$ and $K_s$, respectively. The collective angular momentum provides the other part, which
increases with $I$. Whereas at  $I=7$ the two partner bands (blue and red) look different, they become
 very similar with increasing $I$. The $K_i$ distributions are peaked at $K_i=1$ for the yrast band and $K_i=2$ 
 for the partner band. For both bands the $K_i=0$ component is very small. This means that there is no tunneling
 between the upper and lower hemispheres with respect to the s-l plane, which reflects  the fact that the
 two bands are regular $\Delta I=1$ sequences. The further discussion can be restricted to the upper hemisphere $K_i>0$.
 For $I=7$ the distributions of the yrast and partner band contain only one component $K_i$=1 and 2, respectively. This means that
 the probability does not dependent on the conjugate angle $\Phi_i$, which rotates the vector $\vec J$ of the total angular momentum 
 about the i-axis.   In the case of partner band, a second component at $K_i=4$ develops with increasing $I$. The probability 
 function of  the superposition :
 $A\,\exp[2i\Phi]+B\exp[4i\Phi]$ with $AB<0$\footnote 
{The case $ AB>0$ can be disregarded because the  orientation of the $\vec j$ of the proton hole and the neutron particle along
 respective  long and short axes causes a preferred orientation of $\vec J$ at about $\pi/4$ between these axes.}
 has  maxima  at $\Phi=\pi/4,~3\pi/4,~5\pi/4,~7\pi/4$ and minima at $\Phi=0,~\pi/2,~\pi,~3\pi/2$. 
 The appearance of these maxima signals that chirality develops \cite{FN00}. If chiral symmetry is spontaneously broken, the angular momentum vector
 lies outside the three principal planes of the triaxial density. There are four equivalent orientations in the upper hemisphere, two left-handed and
 two right-handed. Tunneling between these four positions restores chiral symmetry for eigenstates of the Hamiltonian, which is localized around 
 these positions. The tunneling decreases with the length of the vector $\vec J$ and chirality becomes more apparent. It is to be noted that yrast band does not
 develop the same signature of chirality and only a slight admixture of  $K_i$=3 is seen in the even-$I$ states. The fact that the yrast band contains only odd
 $K_i$ and the partner band only even $K_i$ explains why the two band cross without mixing and repel each other.
 
In conclusion, in the present study a systematic investigation of the chiral doublet bands
observed in odd-odd $A \sim$ 100 region has been carried out using the recently developed
multi-quasiparticle triaxial projected shell model approach. The purpose of 
the work was to investigate some puzzling observations reported in this 
mass region. It has been demonstrated that the
observed crossing of the yrast and the partner bands in  $^{106}$Ag is due to the
crossing of the two 2-quasiparticle states having different intrinsic configuration. Further,
it is predicted that a similar crossing for the yrast and the partner bands occurs for
 $^{104}$Ag at a higher angular-momentum of I=18.

\end{document}